\documentclass{article}
\usepackage{graphicx}
\usepackage{amsmath}
\usepackage{hyperref}
\usepackage{authblk}
\usepackage[utf8]{inputenc}

\title{Three particle L\'evy HBT from PHENIX
\thanks{Presented by B. Kurgyis at XIV Workshop on Particle Correlations and Femtoscopy, 3-7 June 2019, Dubna, Russian Federation}}
\author[1]{B\'alint Kurgyis \em{for the PHENIX Collaboration}}
\affil[1]{E\"otv\"os Lor\'and University, Hungary\\H-1117 Budapest, P\'azm\'any P. s. 1/A\\e-mail: \em{kurgyisb@caesar.elte.hu}}
\date{\today}

\begin{document}

\maketitle
\begin{abstract}
Bose-Einstein correlations of identical bosons reveal information about the space-time structure of particle emission from the sQGP formed in ultra-relativistic heavy-ion collisions. Previous measurements of two particle correlations have shown that the source can best be described by a symmetric Levy distribution. Here we report on the measurement of three-particle correlations in 0-30\% centrality Au+Au collisions at $\sqrt{s_{_\text{NN}}} = 200$ GeV, and describe them with a Levy type source. This measurement may shed light on hadron creation mechanisms beyond chaotic emission. We measure three particle correlation strength ($\lambda_3$) as a function of pair transverse momentum. This parameter, combined with two-particle correlation strength $\lambda_2$ may reveal the level of chaoticity and coherence in particle production.    
\end{abstract}

\section{Introduction}
The measurement of Bose-Einstein correlations of identical bosons allows one to map out the femtoscopic geometry of the particle emitting source~\cite{Csorgo:1999sj}. It was found that the particle emitting source corresponding to the sQGP formed in heavy ion collisions is not adequately described by a Gaussian
source~\cite{Adler:2006as,Csanad:2005nr,Afanasiev:2007kk,Adam:2015pbc}, which was often assumed to be the case~\cite{Adler:2004rq,Adamczyk:2014mxp}. Data collected by the PHENIX experiment indicates that this source function is best described by a Lévy type source in one dimensional
measurements~\cite{Adare:2017vig}, and some preliminary results  suggest that this is also the case when a three dimensional analysis is performed~\cite{Kurgyis:2018zck}. Thus, we are going to utilize the Lévy source for the description of the measured three-particle correlation functions as well. From two-particle analyses we can infer the size of the homogeneity region of the source and measure the two particle correlation strength, as well as the Lévy exponent~\cite{Adare:2017vig}. In addition, combining these with three-particle measurements we can explore the nature of particle creation mechanisms in the
source~\cite{Csorgo:1994in,Bolz:1992hc}. Here we present preliminary results on three pion correlation measurements performed by the PHENIX experiment in 0-30\% centrality Au+Au collisions at $\sqrt{s_{_\text{NN}}} = 200$ GeV.

\section{Three-particle femtoscopy with Lévy type source}

We assume that the source can be described by a Lévy stable distribution~\cite{Csorgo:2003uv}:
\begin{equation}
    \mathcal{L}(\mathbf{r};\alpha;\mathbf{R})=\frac{1}{(2\pi)^3}\int \mathrm{d}^3q e^{i\mathbf{qr}} e^{-|\mathbf{qR}|^\alpha}.
\end{equation} 
The distribution is characterized by the Lévy scale parameter $\mathbf{R}$ and the Lévy exponent $\alpha$. We mention two special cases: $\alpha=1$ corresponds to Cauchy distribution, and $\alpha=2$ is a Gaussian distribution.

The three-particle correlation function is defined as:
\begin{equation}
    C_3(\mathbf{k}_1,\mathbf{k}_2,\mathbf{k}_3)=\frac{N_3(\mathbf{k}_1,\mathbf{k}_2,\mathbf{k}_3)}{N_1(\mathbf{k}_1)N_1(\mathbf{k}_2)N_1(\mathbf{k}_3)},
\end{equation}
where $N_3$ and $N_1$ are the three particle and the single particle invariant momentum distributions. 
Assuming a Lévy source, and no final state interaction we can write the interaction-free correlation function as:
\begin{align}
C_3^{(0)}(k_{12}, k_{13}, k_{23}) = 1+ \ell_3e^{-0.5(|2k_{12}R|^\alpha+|2k_{13}R|^\alpha+|2k_{23}R|^\alpha)}\nonumber\\
+\ell_2\bigg(e^{|2k_{12}R|^\alpha}+e^{|2k_{13}R|^\alpha}+e^{|2k_{23}R|^\alpha}\bigg).
\end{align}
The newly introduced notations are the following: $k_{ij}=|k_i-k_j|/2$ is the half of the magnitude of the relative momentum of two particles in their longitudinally comoving frame, $\ell_2$ is the two particle correlation strength measured in three particle correlations and $\ell_3$ is the three particle strength parameter.

We measure the correlation functions of same charged pion triplets, therefore we have to take into account the Coulomb repulsion between them. For this we introduce a Coulomb correction factor ($K_3$), and the Coulomb interacting correlation function is then expressed as:
\begin{equation}
    C_3(k_{12}, k_{13}, k_{23})=K_3(k_{12}, k_{13}, k_{23})C_3^{(0)}(k_{12}, k_{13}, k_{23}).
\end{equation}
For the Coulomb correction factor we use a ``generalized Riverside'' method, which means that we approximate it in the following way~\cite{Gangadharan:2015ina}:
\begin{align}
    K_3(k_{12}, k_{13}, k_{23})&=K_2(k_{12})K_2(k_{13})K_2(k_{23}), \text{ where }\\ K_2(k)&=\frac{\int \mathrm{d}^4r S(\mathbf{r,k})|\Psi^{(Coulomb)}_\mathbf{k}(\mathbf{r})|^2}{\int \mathrm{d}^4r S(\mathbf{r,k})|\Psi^{(0)}_\mathbf{k}(\mathbf{r})|^2}.
\end{align}
Additionally, we add a linear background ($\varepsilon$) and a normalization ($N$) for the correlation function, so the fit function used here is the following:
\begin{equation}\label{eq:fitfunc}
    C_3^{(fit)}=N(1+\varepsilon k_{12})(1+\varepsilon k_{13})(1+\varepsilon k_{23})C_3(k_{12}, k_{13}, k_{23}).
\end{equation}
One defines the two- and three-particle correlation strength parameters as the intercepts of the correlation functions:
\begin{align}
    \lambda_2&=C_2(k_{12}\rightarrow0)-1,\\
    \lambda_3&=C_3(k_{12}=k_{13}=k_{23}\rightarrow 0) -1,
\end{align}
and one finds that $\lambda_3 = \ell_3 + 3\ell_2$.
With two- and three-pion correlations we can investigate the pion creation mechanisms~\cite{Csorgo:1998tn}. First, let us consider the possibility of coherent pion creation. We can introduce the fraction of coherent pions:
\begin{equation}
    p_c=\frac{N_\text{coherent}}{N_\text{coherent}+N_\text{incoherent}},
\end{equation}
where $N_\text{coherent}$ is the number of coherently created pions and $N_\text{incoherent}$ is the number of incoherent ones. Secondly, we introduce the core-halo model. Let us note that not all incoherent pions come from the sQGP (core), but there are long lived resonances whose decay into pions yield a contribution (halo). Therefore, we can define the ratio of primordial pions and the number of pions coming from the core ($N_\text{core}$) and from the halo ($N_\text{halo}$):
\begin{equation}
    f_c=\frac{N_\text{core}}{N_\text{core}+N_\text{halo}}.
\end{equation}
We can express the two- and three particle correlation strengths in terms of the above defined quantities:
\begin{align}
    \lambda_2&=f_c^2\left[(1-p_c)^2+2p_c(1-p_c)\right]\\
    \lambda_3&=2f_c^3\left[(1-p_c)^3+3p_c(1-p_c)^2\right]+3f_c^2\left[(1-p_c)^2+2p_c(1-p_c)\right].
\end{align}
This gives a restriction for the value of $\lambda_3$ in a purely core-halo picture: $0\leq \lambda_3 \leq 5$. Moreover, we can introduce a new, core-halo independent parameter:
\begin{equation}
    \kappa_3=\frac{\lambda_3-3\lambda_2}{2\sqrt{\lambda_2^3}}.
\end{equation}
If partial coherence does not play a role ($p_c=0$) then $\kappa_3=1$. Thus, by measuring $\lambda_3$ and $\kappa_3$ we can investigate the role of partial coherence as well as the appropriacy of the core-halo model.

\section{Results}

\begin{figure}
    \centering
    \includegraphics[width=0.7\textwidth]{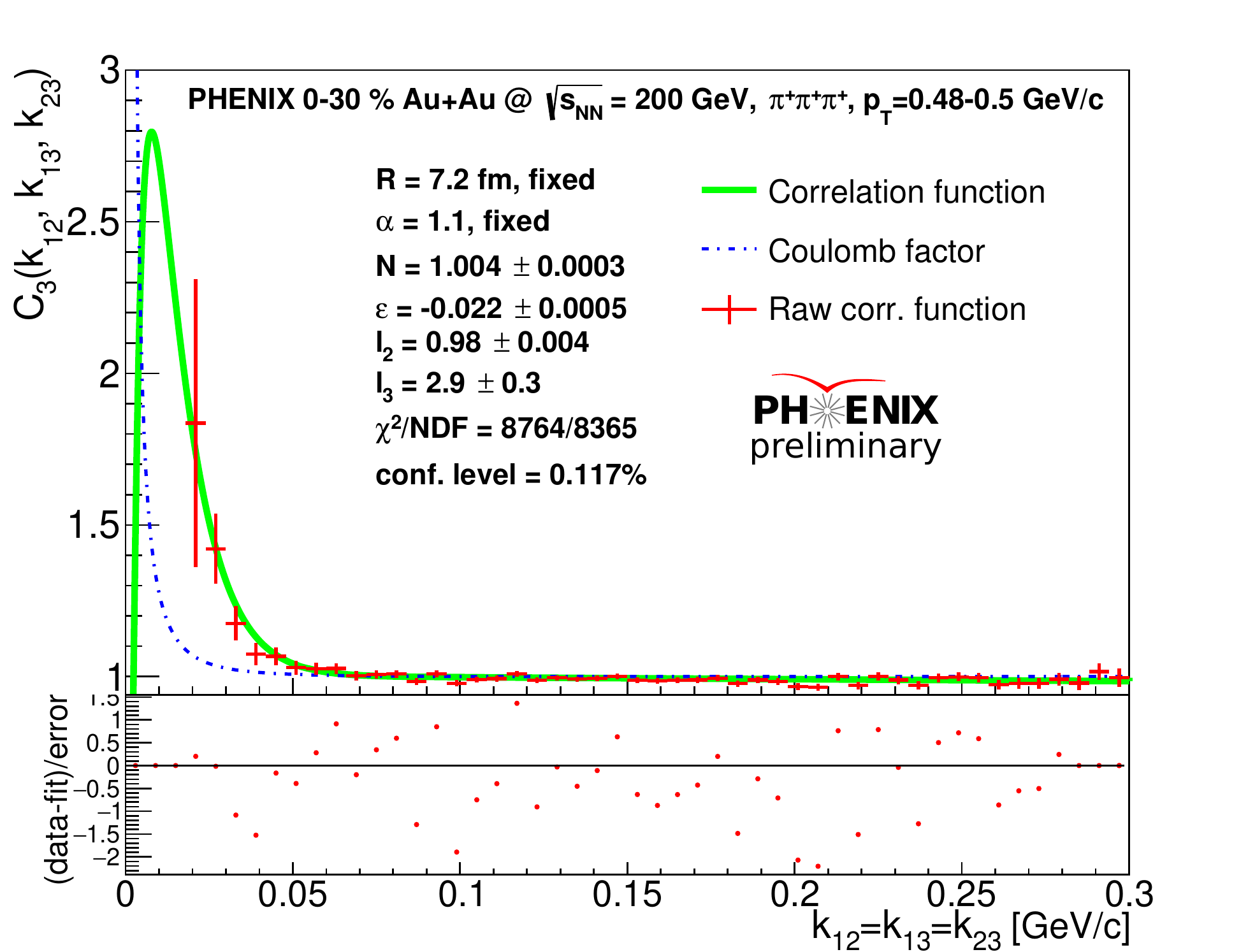}
    \caption{An example plot of the measured correlation function and the fitted function. For visualization we choose to show a diagonal plot.}
    \label{fig:corr_func}
\end{figure}

\begin{figure}
    \centering
    \includegraphics[width=1.0\textwidth]{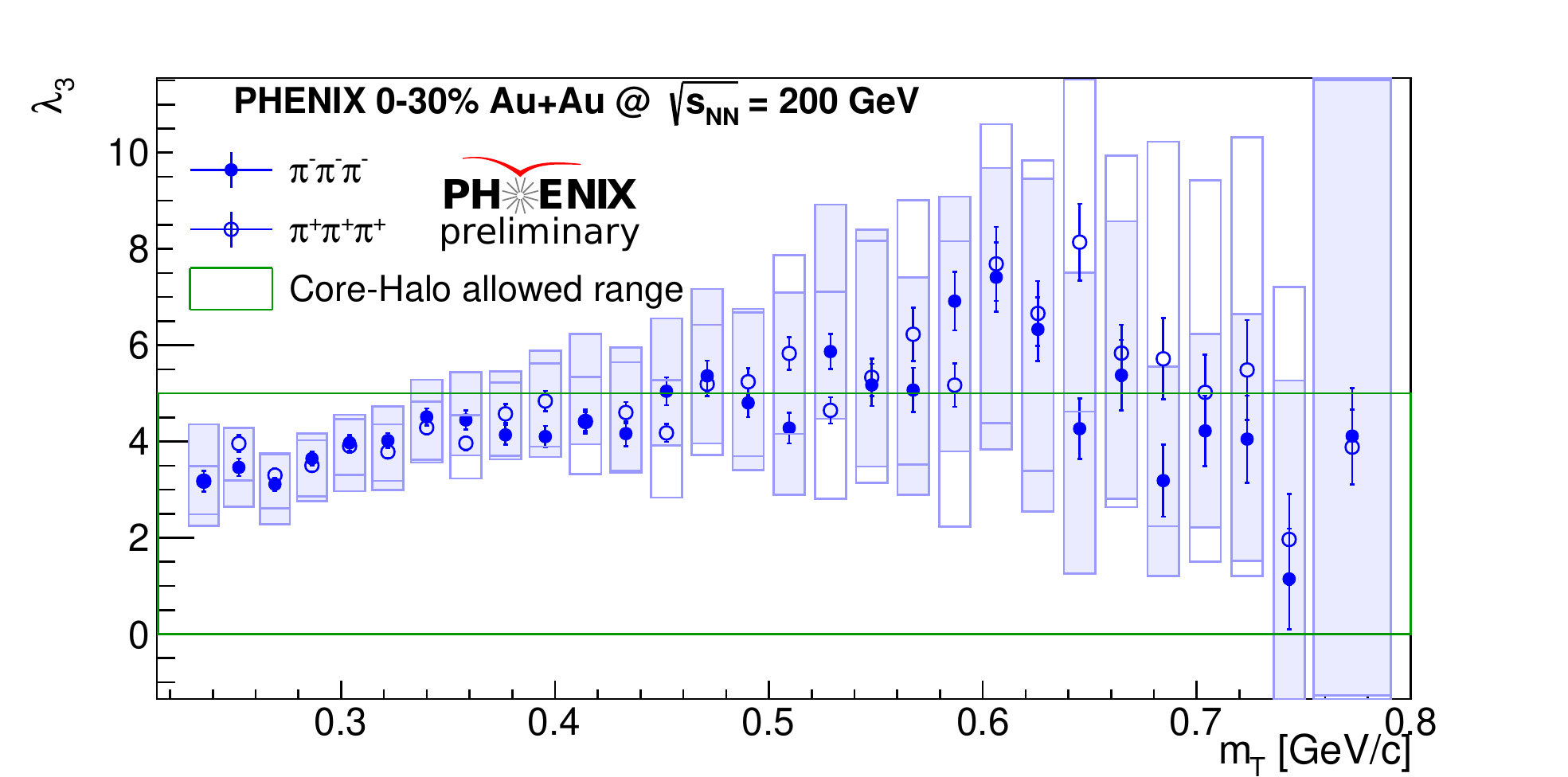}
    \caption{The three particle correlation strength parameter as the function of the transverse mass. The data indicates that we have a system described by a purely core-halo picture as this parameter is in the range of $[0,5]$ within errors.}
    \label{fig:lambda3}
\end{figure}

\begin{figure}
    \centering
    \includegraphics[width=1.0\textwidth]{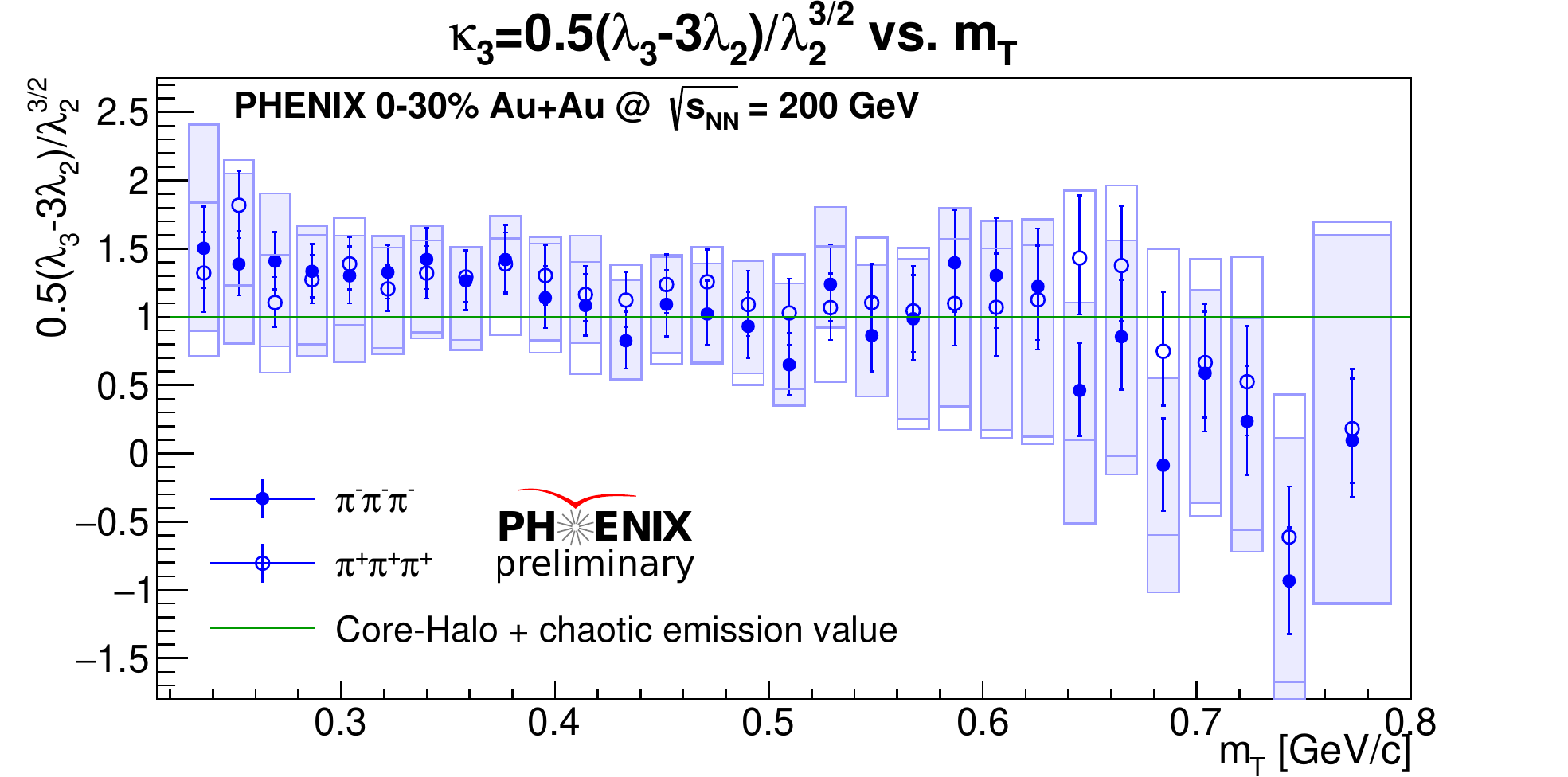}
    \caption{The core-halo independent parameter determined from combining the results of two and three particle correlation measurements. Without partial coherence this parameter should be equal to one, and this holds within errors.}
    \label{fig:kappa3}
\end{figure}

We used $\sqrt{s_{_\text{NN}}}=200$ GeV Au+Au collisions recorded by the PHENIX experiment in the 2010 running period. We restrict ourselves to the 0-30\% most central events. We measure three particle correlation functions of same charged pion triplets for 31 different transverse mass bins and fit the data using the assumption that we are dealing with a Lévy-shaped source. For the visualization of the three dimensional fits, we choose to plot data points and the fitted curve along a diagonal ($k_{12}=k_{13}=k_{23}$) line; this can be seen on Fig.~\ref{fig:corr_func}. For these fits the Lévy scale parameter $R$ and the Lévy exponent $\alpha$ was fixed for the values obtained from two pion correlation measurments done on the same dataset~\cite{Adare:2017vig}. 
The three particle correlation strength, which is calculated from the fit parameters $\ell_2$ and $\ell_3$ is shown as the function of the transverse mass on Fig.~\ref{fig:lambda3}. According to the core-halo model, this parameter has to be between $0$ and $5$ and we can see that this condition is satisfied within errors.
For determining the value of $\kappa_3$ we used the $\lambda_3$ values and the two particle correlation strength values from Ref.~\cite{Adare:2017vig}. The transverse mass dependence of the core-halo independent $\kappa_3$ parameter is shown on Fig.~\ref{fig:kappa3}. Without partial coherence this parameter should be equal to one. We can see, that the preliminary results indicate that $\kappa_3=1$ holds within errors.
\section{Summary}

We have measured three-pion correlation functions in 0-30\% central Au+Au collisions at $\sqrt{s_{_\text{NN}}}=200$ GeV. The results combined with previous two particle measurements enable us to study the core-halo model and the role of partial coherence in the system. Preliminary results on the three particle correlation strength indicate that the system can be well described by a purely chaotic source. Additionally, from the measurement of a core-halo independent parameter we can infer that partial coherence might not play a role in this system, thus the source is chaotic.  
\section*{Acknowledgements}
The author expresses gratitude for the support of Hungarian NKIFH grant No. FK-123842. B. Kurgyis was supported by the UNKP-19-2-II-ELTE-276 New National Excellence Program of the Ministry for Innovation and Technology.

\bibliographystyle{unsrt}

\begin{thebibliography}{99}
\bibitem{Csorgo:1999sj} 
  T.~Cs\"org\H{o},
  Acta Phys.\ Hung.\ A {\bf 15}, 1 (2002)
  [hep-ph/0001233].

\bibitem{Adler:2006as} 
  S.~S.~Adler {\it et al.} [PHENIX Collaboration],
  Phys.\ Rev.\ Lett.\  {\bf 98}, 132301 (2007)
  [nucl-ex/0605032].
 
\bibitem{Csanad:2005nr} 
  M.~Csan\'ad [PHENIX Collaboration],
  Nucl.\ Phys.\ A {\bf 774}, 611 (2006)
  [nucl-ex/0509042].

\bibitem{Afanasiev:2007kk} 
  S.~Afanasiev {\it et al.} [PHENIX Collaboration],
  Phys.\ Rev.\ Lett.\  {\bf 100}, 232301 (2008)
  [arXiv:0712.4372 [nucl-ex]].
  
\bibitem{Adam:2015pbc} 
  J.~Adam {\it et al.} [ALICE Collaboration],
  Phys.\ Rev.\ C {\bf 93}, no. 5, 054908 (2016)
  [arXiv:1512.08902 [nucl-ex]].  
  
\bibitem{Adler:2004rq} 
  S.~S.~Adler {\it et al.} [PHENIX Collaboration],
  Phys.\ Rev.\ Lett.\  {\bf 93}, 152302 (2004)
  [nucl-ex/0401003].
 
\bibitem{Adamczyk:2014mxp} 
  L.~Adamczyk {\it et al.} [STAR Collaboration],
  Phys.\ Rev.\ C {\bf 92}, no. 1, 014904 (2015)
  [arXiv:1403.4972 [nucl-ex]].
  
 
\bibitem{Adare:2017vig} 
  A.~Adare {\it et al.} [PHENIX Collaboration],
  Phys.\ Rev.\ C {\bf 97}, no. 6, 064911 (2018)
  [arXiv:1709.05649 [nucl-ex]].
  
\bibitem{Kurgyis:2018zck} 
  B.~Kurgyis [PHENIX Collaboration],
  Acta Phys. Pol. B Proc. Suppl. vol. 12 (2), pp. 477 - 482 (2019)
  [arXiv:1809.09392 [nucl-ex]].


\bibitem{Csorgo:1994in} 
  T.~Cs\"org\H{o}, B.~L\"orstad and J.~Zim\'anyi,
  Z.\ Phys.\ C {\bf 71}, 491 (1996)
  [hep-ph/9411307].

\bibitem{Bolz:1992hc} 
  J.~Bolz {\it et al.},
  Phys.\ Rev.\ D {\bf 47}, 3860 (1993).
  
\bibitem{Csorgo:2003uv} 
  T.~Cs\"org\H{o}, S.~Hegyi and W.~A.~Zajc,
  Eur.\ Phys.\ J.\ C {\bf 36}, 67 (2004)
  [nucl-th/0310042].
  
\bibitem{Gangadharan:2015ina} 
  D.~Gangadharan,
  Phys.\ Rev.\ C {\bf 92}, no. 1, 014902 (2015)
  [arXiv:1502.02121 [nucl-th]].
  
\bibitem{Csorgo:1998tn} 
  T.~Cs\"org\H{o}, B.~L\"orstad, J.~Schmid-Sorensen and A.~Ster,
  Eur.\ Phys.\ J.\ C {\bf 9}, 275 (1999)
  [hep-ph/9812422].
  
\end{thebibliography}

\end{document}